\documentclass[epj,nopacs]{svjour}
\usepackage{axodraw}
\usepackage{epsfig}
\usepackage{amsfonts}
\usepackage{amsmath}
\usepackage{bbm}
\allowdisplaybreaks[1]

\input pix.sty

\renewcommand{\eq}{eq.~}
\renewcommand{\eqs}{eqs.~}
\renewcommand{\se}{sec.~}

\renewcommand{\fig}{fig.~}

\newcommand{\tinymsbar}{{\overline{\mbox{\tiny\rm{MS}}}}}
\newcommand{\Lambdamsbar}{{\Lambda_\tinymsbar}}

\newcommand{\alphas}{\alpha_{\rm s}}
\newcommand{\Nf}{N_{\rm f}}
\newcommand{\Nc}{N_{\rm c}}
\newcommand{\CF}{C^{ }_\rmii{F}}

\def\lsi{\raise0.3ex\hbox{$<$\kern-0.75em\raise-1.1ex\hbox{$\sim$}}}
\def\gsi{\raise0.3ex\hbox{$>$\kern-0.75em\raise-1.1ex\hbox{$\sim$}}}
\newcommand{\lsim}{\mathop{\lsi}}
\newcommand{\gsim}{\mathop{\gsi}}

\newcommand{\nF}{n_\rmii{F}}
\newcommand{\nB}{n_\rmii{B}}
 \renewcommand{\nF}[1]{f_\rmii{F{#1}}} 
 \renewcommand{\nB}[1]{f_\rmii{B{#1}}} 

\newcommand{\rmii}[1]{{\mbox{\tiny\rm{#1}}}}

\newcommand{\re}{\mathop{\mbox{Re}}}

\newcommand{\Tint}[1]{{\hbox{$\sum$}\!\!\!\!\!\!\!\int\,}_{\!\!\!\!\raise-0.9ex\hbox{$\scriptstyle{#1}$}}}
\newcommand{\Tinti}[1]{{{\Sigma}\!\!\!\!\raise0.3ex\hbox{$\int$}_\rmii{${#1}$}}}

\newcommand{\bi}{\begin{itemize}}
\newcommand{\ei}{\end{itemize}}
\newcommand{\hide}[1]{ }

\newcommand{\ff}{\rmi{\sl f\,}}
\newcommand{\nsum}{{\textstyle{\sum}\,}}
\newcommand{\deltabar}{\raise-0.02em\hbox{$\bar{}$}\hspace*{-0.8mm}{\delta}}
\def\TAsc(#1,#2)(#3,#4,#5)%
{\SetWidth{2.0}\CArc(#1,#2)(#3,#4,#5)\SetWidth{1.0}}
\def\Lwidth{3}

\def\TAgl(#1,#2)(#3,#4,#5){\SetWidth{2.0}\PhotonArc(#1,#2)(#3,#4,#5){\Lwidth}%
{6.283 #3 mul 360 div #4 #5 sub #4 #5 sub mul sqrt mul Tdensity mul}%
\SetWidth{1.0}}
\def\TLgl(#1,#2)(#3,#4){\SetWidth{2.0}\Photon(#1,#2)(#3,#4){\Lwidth}
{#1 #3 sub #1 #3 sub mul #2 #4 sub #2 #4 sub mul add sqrt Tdensity mul}%
\SetWidth{1.0}}

\renewcommand{\picc}[1]{\;\parbox[c]{60pt}{\begin{picture}(60,30)(0,0)
\SetWidth{1.0}\SetScale{1.3} #1 \end{picture}}\; }
\def\Lwidth{1.3}

\newcommand{\picu}[1]{\;\parbox[c]{40pt}{\begin{picture}(50,30)(0,0)
\SetWidth{1.0}\SetScale{0.8} #1 \end{picture}}\; }

\def\ScatA{\picu{%
 \SetWidth{2.0} 
 \Lqu(10,30)(11,29)%
 \Laqu(10,10)(11,11)%
 \SetWidth{1.0} 
 \Line(-1.5,40)(18.5,20)%
 \Line(0,41.5)(21.5,20)%
 \Line(0,-1.5)(21.5,20)%
 \Line(-1.5,0)(18.5,20)%
 \Photon(21.5,20)(40,20){2}{3}%
 \Photon(40,20)(60,40){2}{4}%
 \Photon(40,20)(60,0){-2}{4}%
}}
\def\ScatB{\picu{%
 \SetWidth{2.0} 
 \Lqu(15,35)(16,34)%
 \Laqu(15,5)(16,6)%
 \SetWidth{1.0} 
 \Line(8.5,40)(21.5,27)%
 \Line(10,41.5)(23.5,28)%
 \Line(10,-1.5)(23.5,12)%
 \Line(8.5,0)(21.5,13)%
 \Line(21.5,13)(21.5,27)%
 \Line(23.5,12)(23.5,28)%
 \Photon(23.5,28)(37,41.5){2}{3}%
 \Photon(23.5,12)(37,-1.5){2}{3}%
}}
\def\ScatC{\picu{%
 \SetWidth{2.0} 
 \Lqu(15,35)(16,34)%
 \Laqu(15,5)(16,6)%
 \SetWidth{1.0} 
 \Line(8.5,40)(21.5,27)%
 \Line(10,41.5)(23.5,28)%
 \Line(10,-1.5)(23.5,12)%
 \Line(8.5,0)(21.5,13)%
 \Line(21.5,13)(21.5,27)%
 \Line(23.5,12)(23.5,28)%
 \Photon(23.5,12)(29.5,18){2}{1.5}%
 \Photon(33.5,22)(53,41.5){2}{4}%
 \Photon(23.5,28)(53,-1.5){2}{6.5}%
}}
\def\ScatD{\picu{%
 \SetWidth{2.0} 
 \Lqu(10,30)(11,29)%
 \Laqu(10,10)(11,11)%
 \SetWidth{1.0} 
 \Line(-1.5,40)(18.5,20)%
 \Line(0,41.5)(21.5,20)%
 \Line(0,-1.5)(21.5,20)%
 \Line(-1.5,0)(18.5,20)%
 \Photon(21.5,20)(40,20){2}{3}%
 \Laqu(40,20)(60,40)%
 \Lqu(40,20)(60,0)%
}}
\makeatletter \@addtoreset{equation}{section} \makeatother

\begin{document}
\flushbottom
\sloppy

\title{
 Charm contribution to bulk viscosity
}

\author{M.~Laine 
  \and 
  Kiyoumars A.~Sohrabi 
 }

\institute{
Institute for Theoretical Physics, 
Albert Einstein Center, University of Bern,  
Sidlerstrasse 5, CH-3012 Bern, Switzerland}

\date{December 2014}

\abstract{%
In the range of temperatures reached in future heavy ion collision 
experiments, hadronic pair annihilations and creations of charm quarks
may take place within the lifetime of the plasma. As a result, 
charm quarks may increase the bulk viscosity affecting the early
stages of hydrodynamic expansion. Assuming thermalization,  
we estimate the charm contribution to bulk viscosity within the same 
effective kinetic theory framework in which the light parton contribution 
has been computed previously. The time scale at which this physics
becomes relevant is related to the width of the transport peak 
associated with the trace anomaly correlator, and is found to be 
$\lsim\, 20$~fm/c for $T\,\gsim\, 600$~MeV. 
}

 
\maketitle


%
\section{Introduction}

Recently concrete thoughts about a possible successor to the LHC have been 
aired~\cite{fcc}. 
One of the issues under discussion is whether there is a case for 
a Heavy Ion Collision program at energies much higher than those achievable
at the LHC~\cite{fcc2}. 
An example of a possible new physics observable is that if 
temperatures up to $\sim 1$~GeV could be reached, even charm quarks might
become a chemically equilibrated part of the Quark-Gluon Plasma. 
As the system cools down, their interactions slow down much faster
than those of the light partons. 
Therefore, we might expect a chemical freeze-out
process within an otherwise thermal medium, akin to that associated with 
some Dark Matter scenarios in cosmology. 

The purpose of the present paper is to investigate the role
of charm quarks within the weak-coupling expansion. More precisely, 
we consider a physical observable, the bulk viscosity, which is intimately
related to chemical equilibration. 

Determining transport coefficients
such as the bulk viscosity is a notoriously 
difficult task even within perturbation theory. Starting from massive
quarks (of mass $M$) and very low temperatures ($T \lsim M/\pi$), 
such that thermal resummations might be expected not to play a role, 
it turns out that a field-theoretic determination of the rate of 
pair annihilation and creation necessitates 
a 3-loop computation~\cite{chem}. The situation becomes worse 
at higher temperatures $T \gsim M/g$, where $g^2 \equiv 4 \pi \alphas$,  
when all-orders resummations are needed  
for obtaining the correct leading-order
result~\cite{jy1}. It has been demonstrated, however, that the same
results can be obtained relatively economically by making use of 
effective kinetic theory~\cite{jy2,eff}, 
which describes quasi-particles 
having temperature-dependent properties.\footnote{%
 Naive kinetic theory, without Hard Thermal Loop (HTL) structures
 or Landau-Pomeranchuk-Migdal (LPM) resummed $1\leftrightarrow 2$ 
 splittings included, has been employed 
 as a qualitative tool for
 estimating QCD transport coefficients since
 decades~\cite{hk}. For example an investigation similar in spirit
 to the present one, save for strange quarks, can be found in ref.~\cite{old2}.
 }  
We adopt this framework in the present paper. 

Our goal is to estimate the charm quark
contribution to bulk viscosity, as well as their chemical equilibration
rate, at temperatures between about 200~MeV and 700~MeV. 
In this regime the charm quarks, with a pole mass $M \sim 1.5$~GeV, can 
be considered dilute and perhaps also kinetically equilibrated, 
as is suggested by experiment~\cite{kinetic1,kinetic2}.
Ideally, 
it would be nice to also consider temperatures in the 
ultrarelativistic regime $T \gsim M/g$ in order to be able to compare
with a previous computation~\cite{zeta}, 
but this is not achieved in the present paper.  
Note that if chemical equilibration were to take place, charm quarks should 
be included in the equation of state, where they would have a substantial 
influence already in the temperature range considered~\cite{eos1}--\cite{eos4} 
(they could be relevant even if not fully 
equilibrated~\cite{eos5}). 

The plan of this paper is the following. 
After outlining the general kinetic theory approach in \se\ref{se:general}, 
the specific setup relevant for our problem is detailed in~\se\ref{se:setup}. 
Thermally averaged annihilation rates are evaluated in \se\ref{se:elements}.  
A numerical solution of the basic equations 
is presented in~\se\ref{se:solution}, 
whereas~\se\ref{se:concl} offers some conclusions and an outlook.

%
\section{General approach}
\la{se:general}

We start by outlining the general strategy of the analysis in the language
of kinetic theory and linear algebra. The precise implementation
will be given in \se\ref{se:setup}. Our notation and 
basic philosophy follow closely those presented in ref.~\cite{zeta}, 
to which we refer for further details.

Suppressing all indices, let $S$ be a source vector enforcing
a desired ``adiabatic'' deviation from equilibrium.\footnote{%
 Local equilibrium is to be maintained, but there 
 should be a flow velocity $\vec{u}$
 with $\nabla\cdot\vec{u}\neq 0$.
 } 
The system responds, as dictated
by a linearized collision matrix $C$ ($C^T = C$), 
by deviating the phase space
distributions by an amount $\chi$ around their equilibrium values:
\be
 S = C \chi 
 \;.  \la{SeqCchi}
\ee
The bulk viscosity $\zeta$ is given by the projection of $\chi$
in the direction of $S$: 
\be
 \zeta = S^T \chi \la{zetaeqSchi}
 \;. 
\ee
Let us assume that we know the normalized eigenfunctions $v_n$ of $C$
($v_m^T v^{ }_n = \delta^{ }_{mn}$) and the corresponding eigenvalues
$\lambda_n \in \mathbbm{R}$. 

In general, the collision matrix has zero modes ($C\xi=0$), 
and is therefore not invertible. However, if the system can 
equilibrate, as is expected, then there should be 
a non-trivial solution $\chi\neq 0$ to $S = C \chi$. 
In this case the zero modes are necessarily 
orthogonal to $S$:
 $S^T\xi = \chi^T C^T \xi = \chi^T C \xi = 0$.
Therefore they do not contribute to $\zeta$, and can be omitted. 
In other words, we can restrict to a subspace orthogonal to 
the zero modes, and then invert $C$ in a spectral representation: 
$
 C^{-1} = \sum_n v_n v_n^T / \lambda_n 
$.  
Subsequently, inserting $\chi = C^{-1} S$
into \eq\nr{zetaeqSchi}, 
\be
 \zeta = \sum_n \frac{(S^T v_n)^2}{\lambda_n}
 \;. \la{spectral}
\ee
We observe that the dominant contribution to $\zeta$ is given by 
the smallest eigenvalues, provided that the corresponding $v_n$
have a non-vanishing projection in the direction of $S$. 

The physical realization of this picture is the following. For bulk 
viscosity, the source $S$ breaks conformal invariance. 
If QCD is approximated as a weakly coupled theory with $\Nf = 3$
massless flavours and the massive charm quark, then $S$ is non-zero
only because of the presence of the charm quark, and is
proportional to $e^{-M/T}$. 
Elastic scatterings as well as $1\to 2$ splittings of the light
partons produced in charm decays have rates much faster than 
the charm pair annihilations and creations. 
It is the latter processes which lead to the smallest 
eigenvalues and are therefore considered in the following. 

%
\section{Specific setup}
\la{se:setup}

Bulk viscosity can be defined as a transport coefficient
corresponding to the operator representing (minus) the trace anomaly, 
\be
 \fr13 \Bigl( \sum_{i=1}^{3} T_{ }^{ii} - T_{ }^{00} \Bigr)
 \;, \la{trace}
\ee
where $T^{\mu\nu}$ is the energy-momentum tensor. The part $T^{00}_{ }$
does not contribute to the bulk viscosity, given that its space
average is exactly conserved. It can therefore be left out, 
or subtracted with a suitable coefficient. Within effective kinetic 
theory~\cite{eff}, there is a structure related
to \eq\nr{trace}~\cite{jy2,zeta}, 
representing a shift in pressure ($p = T_{ }^{ii}$) 
when subtracting the contribution of pressure
perturbations carrying energy density 
($(\partial p / \partial e)\delta e \equiv v_s^2 \delta e$, 
where $v_s^2$ is the sound speed squared). For 
a momentum mode $\vec{k}$ and a particle of species $a$, the relevant
subtracted perturbation is of the form~\cite{zeta}
\ba
 q^a(k) & \equiv & 
 \fr13 \vec{v}_k^a \cdot \vec{k} - v_s^2 
 \frac{\partial (\beta E^a_k)}{\partial\beta}
 \;. \la{q_a}
\ea
Here  
$
 \vec{v}^a_k = \nabla^{ }_\vec{k} E^a_k
$ is a group velocity; 
$k \equiv |\vec{k}|$; 
$\beta \equiv 1/T$; 
and $E^a_k$ is an on-shell energy. 
The associated source vector will be denoted by
\ba
 S^a({k}) & \equiv & - T f^a (1\pm f^a)\, q^a(k) 
 \;, \la{S_a}
\ea
where $f^a$ is the equilibrium (Bose or Fermi) distribution
of a particle of type $a$; 
and $\pm$ correspond to bosons and 
fermions, respectively. 

For $v_s^2 = 1/3$ and massless particles
with $E^a_k = k$, 
$S^a$ vanishes identically. More generally, as
pointed out in ref.~\cite{zeta}, all $S^a$'s vanish
identically in a conformal theory. However, with massive quarks
$S^a$ is non-zero even at leading order, as we now show.  

In free SU($\Nc$) gauge theory with $\Nf$ massless flavours and
one massive quark, $v_s^2$ evaluates to  
\be
 v_s^2 = \frac{1}{3} \times \frac{
 {\pi^2 T^5 (8 \CF + 7 \Nf)}  + {60  
 \int_{\vec{q}} q^2 f^{ }_\rmii{F} (1 - f^{ }_\rmii{F})}
  }
 {{\pi^2 T^5 (8 \CF + 7 \Nf)} + {60 
 \int_{\vec{q}} E_q^2 f^{ }_\rmii{F} (1 - f^{ }_\rmii{F})}
  }
 \;, 
\ee
where 
$
 \CF \equiv (\Nc^2-1)/(2\Nc)
$
and
$
 \int_{\vec{q}} \equiv \int\! 
 \frac{{\rm d}^3\vec{q}}{(2\pi)^3}
$.  
Both for $M \ll \pi T$ and $M \gg \pi T$ this reduces to $\fr13$. 
For $M \gsim \pi T$, 
the corrections are exponentially suppressed:
\be
 v_s^2 \approx \fr13 - \frac{20 M^2  \int_{\vec{q}} e^{-E_q/T}}
 {\pi^2 T^5 (8 \CF + 7 \Nf)}
 \;.
\ee
In this case the source vectors of \eq\nr{S_a} read, 
for charm quarks
($S^c$), gluons ($S^g$) and light quarks ($S^q$), respectively,  
\ba
 S^c(k) & \approx & \frac{TM^2}{3 E_k}\, e^{-E_k/T} 
 \;, \la{Sc} \\
 S^g(k) & \approx & - k \, \nB{}(k) \bigl[ 1 + \nB{}(k) \bigr]  
 \,  \frac{20 M^2  \int_{\vec{q}} e^{-E_q/T}}
 {\pi^2 T^4 (8 \CF + 7 \Nf)} 
 \;, \la{Sg} \\ 
 S^q(k) & \approx & - k \,  \nF{}(k) \bigl[ 1 - \nF{}(k) \bigr]  
 \,  \frac{20 M^2 \int_{\vec{q}} e^{-E_q/T}}
 {\pi^2 T^4 (8 \CF + 7 \Nf)} 
 \;.\hspace*{5mm} \la{Sq}
\ea

Let us elaborate on the nature of the approximation that led to 
\eqs\nr{Sc}--\nr{Sq}. 
 It is well known that when considering massive 
 particles at finite temperature, two different types of thermal
 effects appear: exponential corrections and power corrections.
 As an example, consider the integral $\int_\vec{k} (k^m / E_k) \nB{}(E_k)$.
 The Bose distribution can be expanded as
 $\nB{}(E_k) = \sum_{n=1}^{\infty} e^{- n E_k / T}$.
 Given that $e^{-E_k/T} < 1$, this series is absolutely convergent.
 The individual terms in the series lead to 
 \be
   \int_\vec{k} \frac{k^m \, e^{- { n E_k} / { T} } }{ E_k}
  = 
   \frac{
  \Gamma\bigl(\frac{3+m}{2}\bigr)
     }{2\pi^{5/2}} 
   \Bigl(\frac{2MT}{n} \Bigr)^{1+ \frac{m}{2}} 
    K_{1+ \frac{m}{2}}^{ }\Bigl(\frac{nM}{T}\Bigr) 
  \;, 
 \ee 
 where $K$ is a modified Bessel function. 
 The modified Bessel function
 is of the form 
 \be
   K_{1+ \frac{m}{2}}^{ }\Bigl(\frac{nM}{T}\Bigr) \; = \; 
    e^{- {nM} / {T}} 
    \sum_{i=0}^{\infty} c_i\, \Bigl(\frac{T}{nM}\Bigr)^{i + \fr12}
 \;. 
 \ee
 However, the coefficients $c_i$ grow factorially; the 
 series is asymptotic, with a zero radius of convergence.  
 The purpose of the present paper is
 to compute the bulk viscosity to leading order in $e^{-M/T}$ but
 to all orders in $T/M$, avoiding asymptotic series. 
This may be called 
a {\em dilute} approximation, in contrast
to a {\em non-relativistic} approximation which would also truncate
the power corrections to a finite order in $T/M$.

It can easily be verified that the source 
of \eqs\nr{Sc}--\nr{Sq} carries no energy, 
\be
 \sum_{a\, \in \{c,g,q \}} \int_{\vec{k}}  \nu^a S^a(k) E^a_k = 0 
 \;, \la{orthogonal}
\ee
where the degeneracies are $\nu^c \equiv 4\Nc$, 
$\nu^q \equiv 4\Nc\Nf$ and $\nu^g \equiv 2(\Nc^2-1)$. 
This is related to the discussion above \eq\nr{spectral}
concerning zero modes; $E^a_k$ plays the role of $\xi$.
Indeed the full 
collision term of a Boltzmann equation vanishes at any temperature
if equilibrium distributions are used; 
the temperature derivative of this statement asserts that 
$\chi^a(k) \propto E^a_k $ is a zero-mode of
the linearized collision matrix $C$. 
So, \eq\nr{orthogonal} represents a crosscheck that $S^a(k)$
sources a deviation from equilibrium 
in a subspace orthogonal to the zero mode. 

It may be noted that for $T \ll M$, all the terms
in \eqs\nr{Sc}--\nr{Sq} are parametrically of a similar
magnitude $\sim e^{-M/T}$. 
However, the rates at which different particles respond to 
this perturbation are different: number changing rates for 
charm quarks are $\sim \alphas^2 e^{-2M/T}$, whereas for light
partons they are $\gsim \alphas^2$ for $k \ll M$, 
or $\gsim \alphas^2 e^{-M/T}$
for $k \sim M$. Putting the sources $S^a$ on the left-hand side
of the equation (cf.\ \eq\nr{SeqCchi}) and the rates on the right-hand
side, we may estimate the magnitudes of the perturbations $\chi^a$.
We get 
$\chi^c \sim e^{M/T}/\,\alphas^2 \gg \chi^g, \chi^q$. For this reason
$\chi^g,\chi^q$ give an exponentially suppressed contribution 
in \eq\nr{zetaeqSchi}. In other words,  
we can assume light partons to be 
completely thermalized ($\chi^g,\chi^q\sim 0$), and consider
in essence only the block $C^{cc}$ of the 
linearized collision matrix. 

An important further point concerns kinetic equilibration. 
The elastic scatterings of charm quarks on  light partons
are $\sim \alphas^2 e^{-M/T}$. These reactions are again
faster than those corresponding to chemical equilibration, so we may 
assume full kinetic equilibrium to be reached. Full kinetic equilibrium
implies that $\chi^c$ is a $k$-independent constant. Even though we 
consider this situation for our results, we do show $k$-dependence
in our equations for the moment, in order to take the 
limit of a constant $\chi^c$ in a correct way through 
a consideration to be carried out in \se\ref{se:solution}.

For later reference, we add an additional oscillatory
time dependence to the problem, characterized
by a frequency $\omega\neq 0$. 
Then \eq\nr{SeqCchi} becomes~\cite{robert,saremi} 
\be
 S^a(k_1) =  -i \omega f^a (1\pm f^a) \chi^a(k_1;\omega)\, + \! 
 \int_{\vec{k}_2}\!\!\! C^{ab}(k_1;k_2) \chi^b(k_2;\omega)
 \;, \la{source}
\ee 
and the bulk viscosity is obtained as
\be
 \zeta  =  \lim_{\omega\to 0} \zeta(\omega) \;, \quad 
 \zeta(\omega) \; \equiv \; \frac{ 4\Nc }{ T^3 }
 \int_{\vec{k}_1}\!\!\! 
  S^c(k_1) \re \bigl[ \chi^c(k_1;\omega) \bigr]
 \;.  \la{zeta_eff}
\ee

%
\section{Scattering matrix elements}
\la{se:elements}

%
\begin{figure}[t]
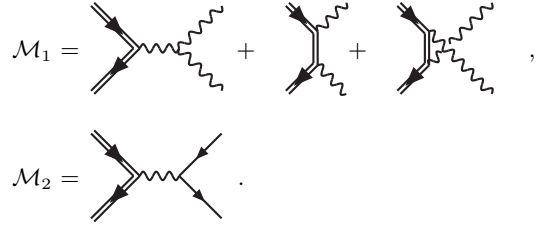


\hspace*{-1cm}
\begin{minipage}[c]{10cm}
\begin{eqnarray*}
 \mathcal{M}^{ }_1 
 & = & 
 \ScatA \quad + 
 \ScatB \hspace*{-5mm} +  
 \ScatC \hspace*{4mm} \;, 
 \\[5mm] 
 \mathcal{M}^{ }_2 
 & = &  
 \ScatD \quad\;. 
\end{eqnarray*}
\end{minipage}

\vspace*{2mm}

\caption[a]{\small 
Processes for pair annihilation or pair creation of charm quarks. 
A double line indicates heavy quarks, a single line light 
quarks, and a wiggly line gluons.}
\la{fig:heur}
\end{figure}
%

In order to solve \eq\nr{source}, 
we need to determine the linearized
collision matrix $C^{cc}$. 
The relevant processes are shown in \fig\ref{fig:heur}.
Summing over all indices, the well-known expressions for the matrix elements
squared read (cf.\ e.g.\ refs.~\cite{gor,pdg})
\ba
 \nsum |\mathcal{M}^{ }_1|^2 
 & = &  
 4 g^4 \CF \Nc \biggl\{ 
 4\Nc\, \frac{(M^2-t)(M^2-u)}{s^2}  
 \nn 
 & + & 
 (2 \CF - \Nc ) 
 \frac{2M^2(s-4M^2)}{(M^2 - t)(M^2 - u)}
 \nn 
 & + & 
 2 \CF \biggl[ 
   \frac{(M^2-t)(M^2-u) - 2 M^2 (M^2+t)}{(M^2-t)^2}
 \nn & & 
   \; + \, 
   \frac{(M^2-t)(M^2-u) - 2 M^2 (M^2+u)}{(M^2-u)^2}
 \biggr]
 \nn 
 & - & 
 2 \Nc 
 \biggl[ 
   \frac{(M^2 - t)(M^2 - u) + M^2(u-t)}{s(M^2 - t)}
 \nn & & 
   \; + \, 
   \frac{(M^2 - t)(M^2 - u) + M^2(t-u)}{s(M^2 - u)}
 \biggr] 
 \biggr\} 
 \;, \hspace*{5mm} \la{M1} \\
 \nsum |\mathcal{M}^{ }_2|^2
  & = & 
 \frac{4 g^4 \CF \Nc \Nf}{s^2}
  \nn & \times & 
 \Bigl[
  (M^2 -t)^2 + (M^2 - u)^2 + 2 M^2 s 
 \Bigr] 
 \;, \la{M2aver}
\ea
where $s,t,u$ are the standard
kinematic variables. 
The bulk viscosity is given by \eq\nr{zeta_eff}, 
where $\chi^c(k_1;\omega)$ is determined 
(up to a discussion concerning kinetic equilibration, 
cf.\ \se\ref{se:solution})
from \eq\nr{source}, {\it viz.}\
\ba
 S^c(k_1) & = &   -i \omega f^{ }_\rmii{F}(E^{ }_{k_1})
 \bigl[ 1 - f^{ }_\rmii{F}(E^{ }_{k_1}) \bigr] \chi^c(k_1;\omega) 
 \nn 
 && + \,
 \int_{\vec{k}_2,\vec{p}_1,\vec{p}_2} \!\!\!
 \frac{ 
 (2\pi)^4 \delta^{(4)}(\mathcal{K}_1 + \mathcal{K}_2 - 
 \mathcal{P}_1 - \mathcal{P}_2) 
 }{16 E^{ }_{k_1} E^{ }_{k_2}\, p_1 \, p_2 }
 \nn 
 && \times \,
 \frac{  f^{ }_\rmii{F}(E^{ }_{k_1})
 f^{ }_\rmii{F}(E^{ }_{k_2})
 }{2\Nc}
 \bigl[ \chi^c(k_1;\omega) + \chi^c(k_2;\omega) \bigr]
 \nn 
 && \times \,
 \biggl\{
   \fr12 \nsum |\mathcal{M}_1|^2  
 \bigl[ 1 + f^{ }_\rmii{B}(p_1) \bigr]
 \bigl[ 1 + f^{ }_\rmii{B}(p_2) \bigr] 
 \nn 
 && \;
  + \, \nsum |\mathcal{M}_2|^2
 \bigl[ 1 - f^{ }_\rmii{F}(p_1) \bigr]
 \bigl[ 1 - f^{ }_\rmii{F}(p_2) \bigr]
 \biggr\} 
 \;. \hspace*{5mm} \la{Boltzmann_2}
\ea
Here $\mathcal{K}_i \equiv (E^{ }_{k_i},\vec{k}_i)$, 
$E^{ }_{k_i}\equiv \sqrt{k_i^2 + M^2}$, 
$\mathcal{P}_i \equiv (p_i,\vec{p}_i)$. Recalling \eq\nr{Sc}, and 
noting that because of energy conservation, $\nB{}(p_i)$
and $\nF{}(p_i)$ are exponentially suppressed, the basic
equation can be reduced to 
\ba
 & & \frac{T M^2}{3 E^{ }_{k_1}} e^{-E^{ }_{k_1}/T}  =   
  -i \omega  e^{-E^{ }_{k_1}/T} \chi^c(k_1;\omega)
 \nn & + &  
 \int_{\vec{k}_2} 
 \frac{e^{-(E^{ }_{k_1} + E^{ }_{k_2})/T}}{4 E^{ }_{k_1} E^{ }_{k_2}}
 \Bigl[ \chi^c(k_1;\omega) + \chi^c(k_2;\omega) \Bigr] \, \Phi(k_1,k_2)
 \;, \nn \la{eq_eff} 
\ea
where we have defined
an average over the angles between $\vec{k}_1$ and $\vec{k}_2$ as
\ba
 \Phi(k_1,k_2) & \equiv & 
 \int\frac{{\rm d}\Omega^{ }_{\vec{k}_1,\vec{k}_2} }{4\pi}
 \nn & \times & 
 \int_{\vec{p}_1,\vec{p}_2} \!\!\!
 \frac{ 
 (2\pi)^4 \delta^{(4)}(\mathcal{K}_1 + \mathcal{K}_2 - 
 \mathcal{P}_1 - \mathcal{P}_2) 
 }{4 p_1 \, p_2 }
 \nn & \times & 
 \frac{ 
   \fr12 \nsum |\mathcal{M}_1|^2  
  + \nsum |\mathcal{M}_2|^2
  }{2\Nc}
 \;. \la{Phi_def}
\ea

The integrals in \eq\nr{Phi_def} can be carried out in closed form. 
Writing 
$
  \Phi(k_1,k_2) = 
   \Phi^{ }_\rmi{b}(k_1,k_2)
 + 
   \Phi^{ }_\rmi{f}(k_1,k_2)  
$,
where $\Phi^{ }_\rmi{b}$ originates from the bosonic final states
(the amplitude $\mathcal{M}^{ }_1$ in \fig\ref{fig:heur}) and 
$\Phi^{ }_\rmi{f}$ from the fermionic ones
(the amplitude $\mathcal{M}^{ }_2$),
the fermionic part reads
\be
 \Phi^{ }_\rmi{f}(k_1,k_2) =   \frac{g^4 \CF \Nf}{6\pi}
 \biggl\{
   1
   -  \frac{M^2}{2 k_1 k_2}
  \ln 
  \biggl(  
 \frac{\Delta^{ }_{+-}} 
      {\Delta^{ }_{++}} 
  \biggr)
 \biggr\}
 \;,  \la{Phii_f}
\ee
where
\be
 \Delta^{ }_{\sigma\tau} \equiv 
 E^{ }_{k_1} E^{ }_{k_2} + \sigma\, M^2 + \tau\, k_1 \, k_2 
 \;.
\ee
For $k_1,k_2\to 0$ this reduces to 
$
 \Phi^{ }_\rmi{f}(0,0) = { g^4 \CF \Nf } / { (4\pi) }
$.
In the bosonic case we get 
\ba
 && \hspace*{-0.5cm} 
 \frac{2\pi\, \Phi^{ }_\rmi{b}(k_1,k_2)}{g^4 \CF}
  \la{Phii_b} \\ 
  & = & 
   \Nc\, \biggl\{
    -\fr13 + \frac{M^2}{k_1k_2}
  \biggl[
  \sqrt{\frac{\Delta^{ }_{-+}} 
             {\Delta^{ }_{++}}} 
  \mathop{\mbox{atanh}} 
  \sqrt{\frac{\Delta^{ }_{-+}} 
             {\Delta^{ }_{++}}} 
 \nn & & \; - \, 
  \sqrt{\frac{\Delta^{ }_{--}} 
             {\Delta^{ }_{+-}}} 
  \mathop{\mbox{atanh}} 
  \sqrt{\frac{\Delta^{ }_{--}} 
             {\Delta^{ }_{+-}}} 
 \; + \, 
 \frac{11}{12} \ln 
 \biggl(  
 \frac{\Delta^{ }_{+-}} 
      {\Delta^{ }_{++}} 
  \biggr)
      \biggr]
 \biggr\}
 \nn & + & 
   \CF\, \biggl\{
    -2 + \frac{1}{k_1k_2}
  \biggl[
        \frac{\Delta^{3/2}_{-+}}
             {\Delta^{1/2}_{++}}
  \mathop{\mbox{atanh}} 
  \sqrt{\frac{\Delta^{ }_{-+}} 
             {\Delta^{ }_{++}}} 
 \nn & &
  \; - \, 
   \frac{\Delta^{3/2}_{--}}
         {\Delta^{1/2}_{+-}} 
  \mathop{\mbox{atanh}} 
  \sqrt{\frac{\Delta^{ }_{--}} 
             {\Delta^{ }_{+-}}} 
 \nn & & \; + \, 
 3M^2   
 \biggl(   
 \mathop{\mbox{atanh}^2} 
 \sqrt{\frac{\Delta^{ }_{-+}} 
             {\Delta^{ }_{++}}} 
 - 
 \mathop{\mbox{atanh}^2} 
 \sqrt{\frac{\Delta^{ }_{--}} 
             {\Delta^{ }_{+-}}} 
  \biggr)
      \biggr]
 \biggr\} 
 \;. \nonumber
\ea
For $k_1,k_2\to 0$, we get
$
 \Phi^{ }_\rmi{b}(0,0) = {g^4 \CF}
 \bigl( 2\CF - \frac{\Nc}{2} \bigr) / ({4\pi})
$.

In the extreme limit $T \ll M/\pi$, the integrals in \eqs\nr{zeta_eff}, 
\nr{eq_eff} are saturated by $k_1,k_2 \sim \sqrt{2TM} \ll M$. Then we may
invoke an approximation in which the energies appearing in the exponentials
are set to their non-relativistic forms, $E^{ }_{k_i} \approx M + k_i^2/(2M)$,
and the energies in the prefactors are expanded to first non-trivial order
in $k_1, k_2$. 
This leads to power-suppressed
thermal corrections. 
As mentioned in the paragraph following \eq\nr{Sq}, 
in our main results (\se\ref{se:solution}) we avoid
making such an approximation because of its 
questionable convergence. However, for future reference, 
let us work out
the limiting values in the remainder of the present section. 

Setting $k_1,k_2\to 0$ where their effects are subleading; making use of 
the expressions for $\Phi^{ }_\rmi{f}(0,0)$ and 
$\Phi^{ }_\rmi{b}(0,0)$ mentioned above; 
and carrying out the integral over $\vec{k}_2$, \eq\nr{eq_eff} can be 
solved as 
\be
 \chi^c(0;\omega) = \frac{TM}{3} \frac{1}{- i\omega + \Gamma^{ }_\rmii{chem}}
 \;, \la{chic0}
\ee 
where we have defined (following ref.~\cite{chem})
\ba
 \Gamma^{ }_\rmi{chem} & \equiv & 
 \frac{1}{2M^2}
 \left( \frac{TM}{2\pi} \right)^{\fr32}
 e^{-M/T}
 \Bigl[
     \Phi^{ }_\rmi{b}(0,0) +  \Phi^{ }_\rmi{f}(0,0)
 \Bigr]
 \nn & = &  
 \frac{g^4 \CF}{8\pi M^2}
 \left( \Nf + 2 \CF - \fr{\Nc}2 \right)
 \left( \frac{TM}{2\pi} \right)^{\fr32}
 e^{-M/T}
 \;. \nn \la{Gamma_heur}
\ea
Inserting \eqs\nr{Sc} and 
\nr{chic0} into \eq\nr{zeta_eff} yields 
\ba
 \zeta 
 & = & 
 \lim_{\omega\to 0} \frac{M^2}{9T} \, 
 \frac{\chi^{ }_{\ff} \Gamma^{ }_\rmii{chem}}
 {\omega^2 + \Gamma^2_\rmii{chem} }
 \nn 
 & = &  
 \frac{32\pi M^4 \Nc}{9 g^4 T \CF 
 (\Nf + 2 \CF - {\Nc} / 2)}
 \;\equiv\; 
 \zeta^{ }_0
 \;, \la{zeta_0}
\ea
where the charm quark susceptibility was defined as 
$
 \chi^{ }_{\ff} \equiv 
 4 \Nc\, \bigl( \frac{T M}{2\pi} \bigr)^{3/2} e^{- M / T}
$.
Eq.~\nr{zeta_0} agrees with an expression given 
in ref.~\cite{bulk}, and is used for normalization below. 

%
\section{Numerical solution}
\la{se:solution}

 The purpose of this section is to present a numerical evaluation
 of the final integrals that are needed for determining the bulk
 viscosity to leading order in $e^{-M/T}$ but to all orders in $T/M$.
 As has been discussed in \se\ref{se:setup}, 
 for theoretically consistent results this is to be done by assuming
 full kinetic equilibrium, whereby the function $\chi^{c}$ is
 constant. However, a constant value needs to be imposed in a correct way. 
 Allowing first for general $k$-dependence, we  define 
 a quadratic form as 
$
 Q[\chi] \equiv  \langle S,\chi \rangle 
 - \fr12  \langle \chi, [C-i\omega \nF{}(1-\nF{})]\chi \rangle
$.
The bulk viscosity is formally given as the extremal value:
\be
 \zeta = 2 \lim_{\omega\to 0}
 \re \left. Q(\chi) \right|_{\delta Q / \delta \chi = 0}  
 \;. \la{extremal}
\ee
At this point we can restrict the function space to that 
of constant functions. 
Then \eq\nr{eq_eff} yields, for the observable defined in \eq\nr{zeta_eff},  
\be
 \zeta(\omega) = \frac{ 4\Nc\, {\tilde{S}}^2\, \tilde{C} }{ 
         {T^3\, ( \tilde{C}}^2 + {\tilde{\omega}}^2 )}
 \;, \la{zw}
\ee
where 
\ba
 \tilde S  & \equiv &  
 \int_\vec{k} 
 \frac{TM^2 e^{-E^{ }_{k}/T}}{3E^{ }_{k}} 
 \;, \\
 \tilde C^{ }  & \equiv &  
 \int_{\vec{k}_1,\vec{k}_2} \!\!  
 \frac{e^{-(E^{ }_{k_1}+E^{ }_{k_2})/T}}
 {2 E^{ }_{k_1} E^{ }_{k_2} }
 \; \Phi(k_1,k_2) \,
 \;,  \\
 \tilde \omega^{ }  & \equiv & 
 \omega \int_{\vec{k}} 
 e^{-E^{ }_{k}/T}
 \;,
\ea
and the weight $\Phi$ is from
\eqs\nr{Phii_f} and \nr{Phii_b}.

In \fig\ref{fig:zeta}, we plot the results for $\zeta$ obtained with 
the above procedure. The physics conclusion 
is that when the temperature increases, 
$\zeta/\zeta^{ }_0$, where $\zeta^{ }_0$ is from \eq\nr{zeta_0}, 
decreases below unity. Our approximation, which assumed $e^{-M/T}$
small, is reliable at most for $T \lsim M/\pi$. 
Nevertheless, it seems conceivable that the 
solution eventually extrapolates to the result obtained 
for $T \gg M \gg \alphas T$ in ref.~\cite{zeta}
(shown with a grey band). 

\begin{figure}[t]


\centerline{%
 \epsfysize=7.5cm\epsfbox{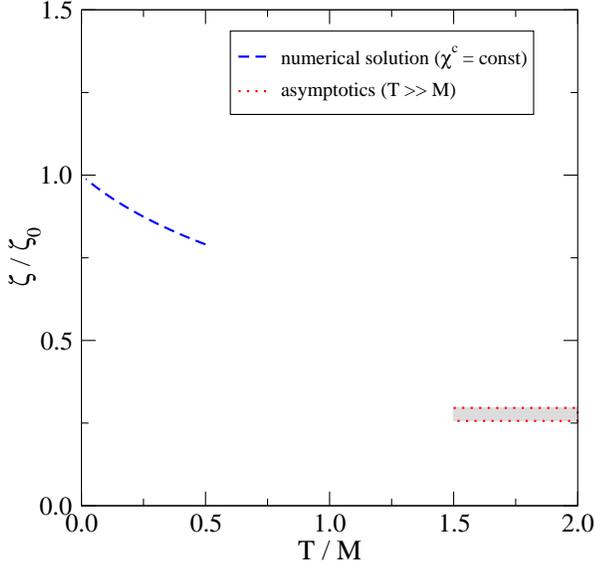}%
}

\caption[a]{\small
The bulk viscosity, normalized to \eq\nr{zeta_0}, 
as a function of $T/M$, for $\Nc = \Nf = 3$. 
We have also compared with the value for 
$T \gg M \gg \alphas T$ from ref.~\cite{zeta}, 
for $\alphas = 0.2 \ldots 0.3$ and $\Nf = 3$.   
}

\la{fig:zeta}
\end{figure}

\begin{figure}[t]


\centerline{%
 \epsfysize=7.5cm\epsfbox{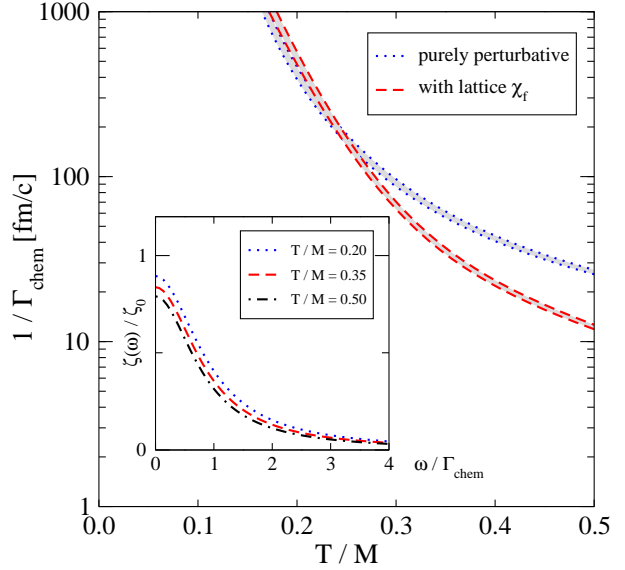}%
}

\caption[a]{\small
The inverse of $\Gamma^{ }_\rmi{chem}$ 
in physical units, for $M = 1.5$~GeV and 
an effective thermal gauge coupling from ref.~\cite{gE}
(we set $\Lambdamsbar \simeq 360$~MeV from ref.~\cite{pacs}; 
this leads to $\alphas\simeq 0.3$ at $T/M \simeq 0.2$
and $\alphas\simeq 0.2$ at $T/M \simeq 0.5$). 
For comparison we also plot the result from 
ref.~\cite{chem} where partial lattice input~\cite{eos4,ding2} 
has been included. The insert shows
the $\omega$-dependence of the transport peak from \eq\nr{zw}.
The axes are normalized to \eqs\nr{Gamma_heur} and \nr{zeta_0}.
}

\la{fig:peak}
\end{figure}

The frequency dependence of the transport peak is plotted in the insert of 
\fig\ref{fig:peak}. Frequency has been normalized to
$\Gamma_\rmi{chem}^{ }$ from \eq\nr{Gamma_heur}; 
$\Gamma_\rmi{chem}^{ }$ is plotted in physical units 
in \fig\ref{fig:peak}. 
The inverse of the width of the Lorentzian
shape serves as an estimate for the microscopic time scale above which the 
charm contribution to the bulk viscosity plays a role within an 
effective hydrodynamic description. If the actual time scale characterizing
the hydrodynamic expansion of the system is 
shorter than the microscopic one, a different approach, with the charm
density appearing as an additional dynamical variable on par
with the temperature and flow velocity, would be needed. 

%
\section{Conclusions}
\la{se:concl}

We have shown that 
in the limit $M \gsim \pi T$, the bulk viscosity of deconfined
QCD plasma grows as $\zeta \simeq 0.04 M^4 / (\alphas^2 T)$, 
cf.\ \eq\nr{zeta_0} and \fig\ref{fig:zeta} which shows
that \eq\nr{zeta_0} provides for a reasonable estimate in this
temperature range. Compared with the 
result $\zeta \simeq 0.25 \alphas^2 T^3$ for $M \ll \alphas T$~\cite{zeta}, 
we note that massive quarks dominate the QCD bulk viscosity if
$M \gsim 1.6 \alphas T$. If the system lived for an infinitely 
long time, this would be true for charm quarks
at all temperatures that can conceivably be reached in current or future
heavy ion collision experiments. However, it is very important
to keep  
in mind that dynamical reactions need to take place in order for this
contribution to the bulk viscosity 
to be physically relevant for the hydrodynamical
evolution of a system with a finite lifetime.\footnote{%
 This requirement is much easier to satisfy in cosmology than in heavy
 ion collision experiments, because the expansion rate of the Universe
 is inversely proportional to the Planck mass and thus much smaller. 
 Correspondingly effects from particle 
 mass thresholds could have a favourable influence e.g.\ 
 on scalar field friction coefficients
 directly proportional to the bulk viscosity~\cite{bulk1,bulk2}.
 } 

The time scale needed for 
the bulk viscosity to ``build up''  
can be estimated from the inverse of the width of the transport peak 
associated with the trace anomaly, and is parametrically of the form
$\tau \simeq 0.45 M^{1/2} e^{M/T} / (\alphas^2 T^{3/2})$, 
cf.\ \eq\nr{Gamma_heur} and \fig\ref{fig:peak} which suggests
that \eq\nr{Gamma_heur} is a conservative estimate in the temperature
range considered.
The associated processes take place 
within the lifetime of the plasma generated in heavy ion 
collision experiments only if a temperature $T \gsim 600$~MeV 
can be reached for an extended period. For $T \lsim 400$~MeV 
charm and anticharm quarks constitute separate conserved charges
like in a statistical model~\cite{stat}. In between, a partial
chemical equilibration may take place, and a dynamical simulation, 
perhaps as a further ingredient in 
an existing one for kinetic equilibration as reviewed in ref.~\cite{ab},  
would be needed for judging the practical importance of charm quarks. 
For a recent study of the practical effects of a bulk viscosity, 
see e.g.\ ref.~\cite{paquet}.

Apart from their effect on bulk viscosity, chemically equilibrated charm
quarks would also be relevant for basic thermodynamic quantities such as the
equation of state~\cite{eos1}--\cite{eos4}. In addition they could change
the shear viscosity by a noticeable amount. 

We end by stressing that obtaining complete leading-order results 
for $T\sim M$
necessitates implementing more complicated resummations than  
have been considered here~\cite{zeta}. 
The same is also true if we decrease the temperature down to 
$T \lsim \alphas^2 M$~\cite{sommerfeld}, which is in principle 
a regime relevant for bottom quarks.  
Beyond resummed computations, 
it would also be interesting
to address the corresponding physics on the lattice. Comparisons
of NLO perturbation theory~\cite{bulk,GStau} 
and lattice simulations~\cite{ding2} for imaginary-time correlators 
suggest that lattice data are already close to the continuum limit, 
so that future progress may be expected.

%
\section*{Acknowledgements}

We are grateful to G.D.~Moore for helpful discussions. 
This work was partly supported by the Swiss National Science Foundation
(SNF) under grant 200020-155935. 
M.L thanks the Institute for Nuclear Theory at the University of Washington 
for hospitality during the completion of this work.

%

\end{document}